\documentclass{iau} 
\usepackage{graphicx}

\title[Quiescent Black Hole Binaries] 
{Black Hole Binaries in Quiescence}

\author 
{Charles D. Bailyn}

\affiliation{Department of Astronomy, Yale University, \\ PO Box 208101, New Haven CT, 06520-8101 USA\\ email: {\tt charles.bailyn@yale.edu} \\[\affilskip]
Yale-NUS College \\ 16 College Avenue West, Singapore, 138527}

\pubyear{2016}
\volume{324}  
\jname{New Frontiers in Black Hole Physics}
\begin{document}

\maketitle

\begin{abstract}

I discuss some of what is known and unknown about the behavior of black hole binary systems in the quiescent accretion state.  Quiescence is important for several reasons: 1) the dominance of the companion star in  optical and IR wavelengths allows the binary parameters to be robustly determined --- as an example, we argue that the longer proposed distance to the X-ray source GRO J1655-40 is correct;  2) quiescence represents the limiting case of an extremely low accretion rate, in which both accretion and jets can be observed; 3) understanding the evolution and duration of the quiescent state is a key factor in determining the overall demographics of X-ray binaries, which has taken on a new importance in the era of gravitational wave astronomy.

\keywords{binaries: close, black hole physics, gravitational waves, X-rays: binaries}
\end{abstract}

\firstsection 
\section{Introduction}

\begin{figure}[b]
 \vspace*{0 cm}
\begin{center}
\hspace{0.2cm}
\includegraphics[width=1.4in]{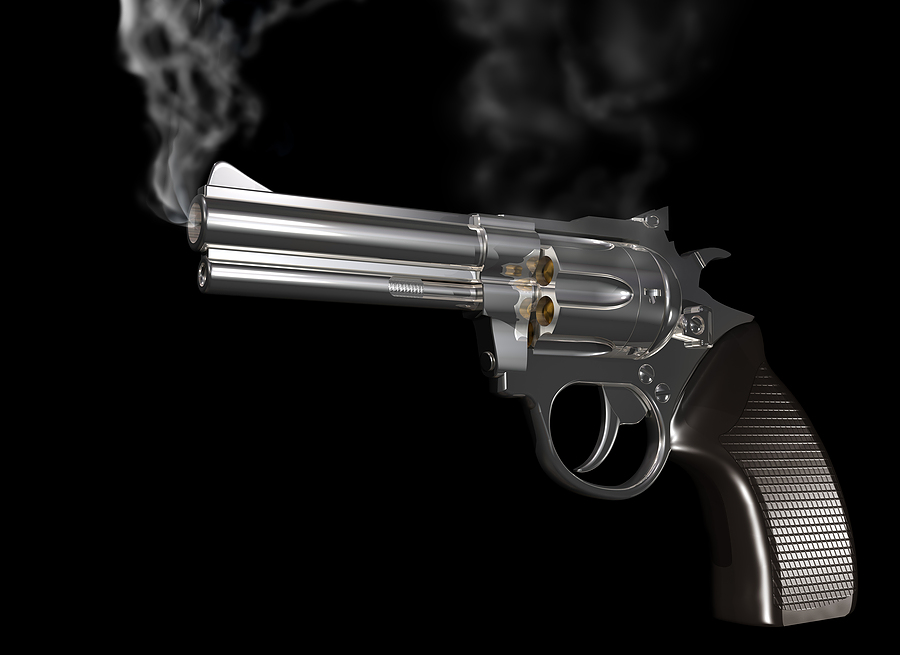}
\includegraphics[width=1.4in]{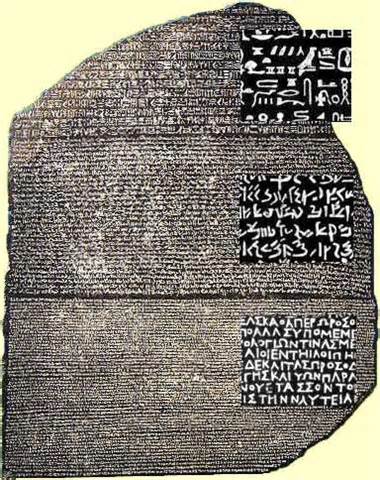}
\includegraphics[width=1.4in]{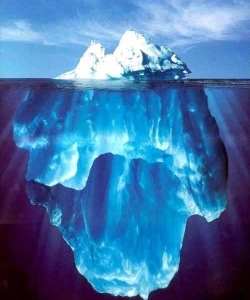}
 \vspace*{0 cm}
 \caption{Three well-known astrophysical metaphors}
  \label{fig1}
\end{center}
\end{figure}

Most black hole X-ray binaries are transients.  They spend years to decades in a very low state, and then increase in luminosity over a few days by as much as eight orders of magnitude.  The outbursts last for weeks to months before the source returns to quiescence.  Quiescent black hole binaries (hereafter QBHBs), have optical/IR emission dominated by the companion star, which can be used to precisely determine the orbital parameters, including the mass of the black hole (see \cite{K12} for details and a discussion of potential errors).  But there is also emission from the residual accretion flow and from jets, which plays an important role in understanding the changes in accretion state during the outburst.  Finally, the inferred demographics of X-ray binaries, which have assumed a heightened importance in the dawning era of gravitational wave astronomy (\cite{A16}), depend critically on the duration of quiescence.  

Observational astrophysicists commonly employ a variety of metaphors in explaining the importance of their work.  Accordingly I will organize the rest of this paper by means of three of the more common metaphors, whose literal manifestations are shown in Figure 1.
I will take as an example the prototypical black hole X-ray transient A0620-00 (hereafter A0620), which underwent a huge outburst in 1975 (\cite{E75}) reaching a peak flux of over 50 Crabs. and has been in quiescence ever since.

\section{The Smoking Gun: Evidence for Black Holes in QBHBs}

A ``Smoking Gun'' is an observation that confirms something that was already suspected to be true.  In the case of QBHBs, the confirmed truth is that the accreting object in many X-ray binaries is a black hole.

The companion stars which dominate the optical/IR light in QBHBs have optical absorption spectra similar to those of ordinary main sequence or giant stars. By applying traditional binary star techniques to these stars, the orbital parameters of the system can be robustly obtained.  In many cases, the mass of the compact accreting star can be shown to be greater than $3M_{\odot }$, the upper limit on the mass of a neutron star.  Such objects are sometimes referred to as ``dynamically confirmed black hole candidates''.

The radial velocity curve of the companion can be used to determine the mass function $M_f = PK^3/(2\pi G)$, which is a strict lower limit on the mass of the compact object.  The first mass function greater than $3M_{\odot }$ was obtained by \cite{MR86} for A0620; since then better data have been obtained for A0620 (\cite{N08}) and over a dozen other systems have been observed (\cite{CS16}).  The radial velocity curves are beautifully sinusoidal, as expected from tidally circularized systems, and there is little doubt that they represent the true motion of the companion star.

To obtain a mass, rather than a mass function, for the compact object, information about the orbital inclination $i$ must be obtained.  This is generally done by studying the ellipsoidal variations of the companion star, caused by the tidally distorted shape of the companion, which results in changes in the cross-sectional area of the star as a function of orbital phase.  The interpretation of such observations requires accounting for the contribution of the accretion flow to the observed flux (e.g \cite{C10} for A0620).
When this is done, the derived binary parameters fix the geometric size of the companion quite precisely, and the observed colors and absorption spectrum fix the temperature.  Therefore the intrinsic luminosity of the companion is well-determined, leading to robust determinations of the distance.   

One relevant case is that of GRO J1655-40, which shows very clearly defined ellipsoidal variations (\cite{G01}, \cite{BP02}). The well-determined physical size and temperature of the companion require a distance of $\gtrsim 3$ kpc. An alternative distance of $\lesssim 1.7$ kpc (\cite{F06}) would require the companion star to underfill its Roche lobe by a factor of 2, which would result in a nearly spherical star that cannot produce the observed ellipsoidal variations.  The lower distance estimate is based on the assumption that the observed spectral type implies a star with mass and radius similar to those of isolated field stars.  But the evolution and surface gravity of a star losing mass through Roche lobe overflow differ dramatically from those of isolated stars, and the observed masses and radii of QBHB companions differ from field stars of similar spectral type in many well-studied systems, including A0620 (\cite{C10}) and V4641 Sgr (\cite{M14}), as well as GRO J1655-40.

\section{The Rosetta Stone: The Disk-Jet Connection}

A ``Rosetta Stone" is an observation that connects two hitherto disparate phenomena.  QBHBs may provide such a connection between accretion physics and jet outflows.

\begin{figure}[b]
 \vspace*{-0.5cm}
\begin{center}
\includegraphics[width=3.5in]{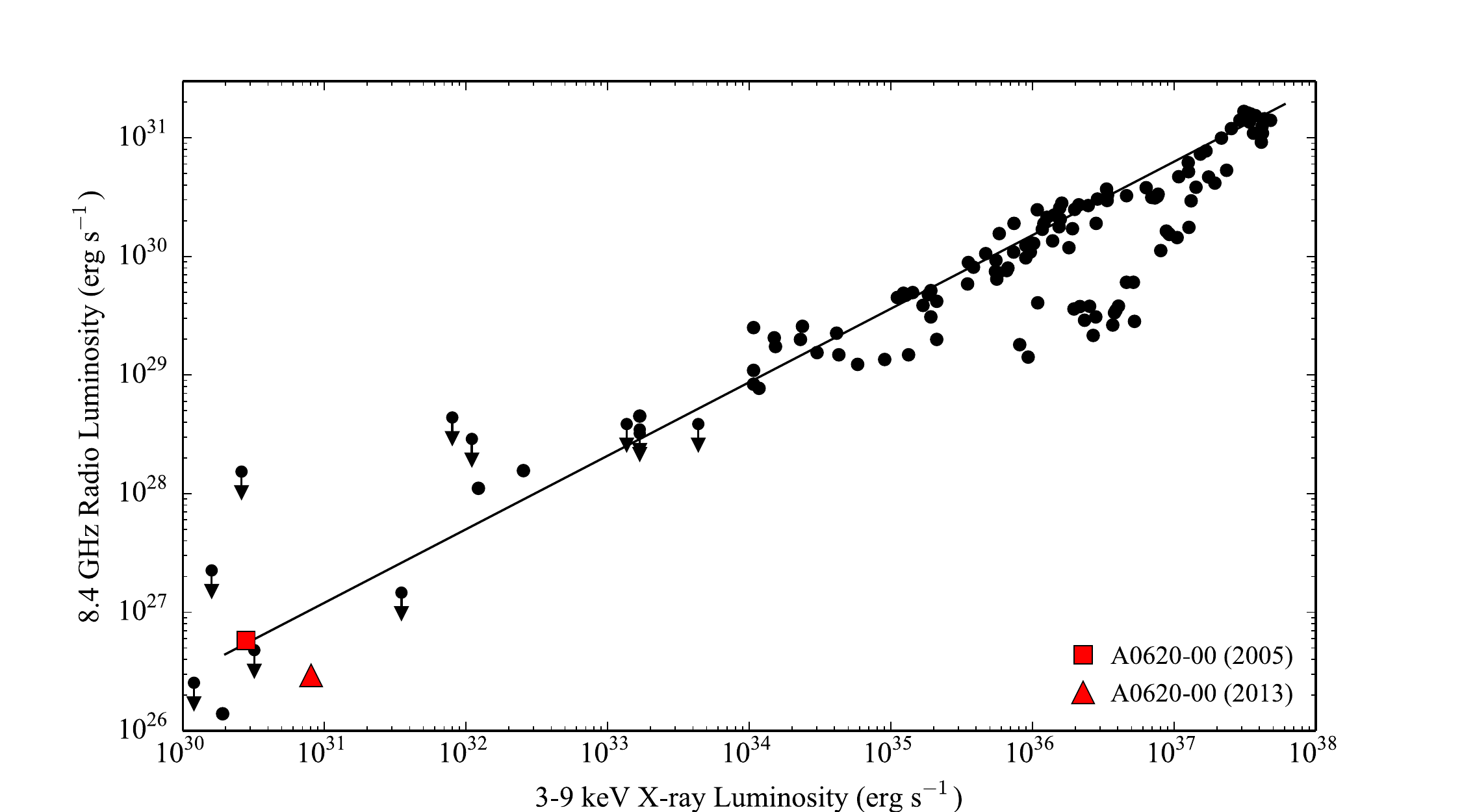} 
 \vspace*{0cm}
 \caption{Radio/X-ray correlation for X-ray binaries --- Din\c{c}er et al. in preparation.  The line represents the X-ray/radio correlation defined by \cite{C13}.}
  \label{fig3}
\end{center}
\end{figure}

During the transition from quiescence to outburst and back, X-ray binaries change not only the amount of accretion, but the nature and state of the accretion flow.  These so-called ``state changes'' have been the subject of considerable recent work.  The general consensus (see \cite{RM06} for a review) is that there is a low-hard state, with a power-law X-ray spectrum and radio emission from a jet, and a
high-soft state, in which the X-ray luminosity is dominated by a thermal accretion disk , which quenches the jet.  There are also a variety of intermediate states.

The presence of a jet in the low-hard state has led to a scenario in which the accretion in that state is in the form of a radiatively inefficient radial flow, which allows the formation of  of a jet (\cite{F03}). But it should be noted that the empirical support for such accretion-jet scenarios rests on the radio/X-ray correlation.  The possibility that there might be a ``universal'' correlation in which $L_R \propto L_X^{0.6}$, and, with suitable correction for the mass of the accretor, extends to the AGN, generated considerable excitement (\cite{C13}). Observations of A0620 (\cite{G06}) held down the low luminosity end of this correlation nicely. However, further investigation revealed multiple correlations and branches, and a recent reobservation of A0620 has shown that the source has moved in the $(L_X,L_R)$ plane in a direction perpendicular to the general correlation (see Figure 2).  So the connection between accretion and jet physics is not yet as empirically well-founded as one might have hoped.

\section{The Tip of the Iceberg: QBHB Demographics}

The ``Tip of the Iceberg'' is an object or observation which implies the existence of a large unseen population.  The current set of observed X-ray transients is certainly the tip of the iceberg of many currently unobserved galactic black hole binaries.

\begin{figure}[b]
 \vspace*{-5.5 cm}
\begin{center}
\includegraphics[width=5in]{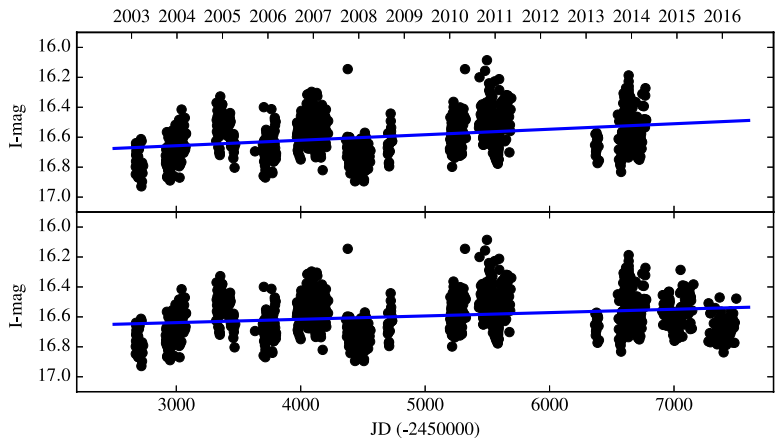} 
 \vspace*{-5.0 cm}
 \caption{Photometry of A0620 from 2003 through 2014 and 2016 with an overall linear fit.  Note the decrease in the slope when the data from 2015-16 are included.}
  \label{fig3}
\end{center}
\end{figure}

X-ray transients are generally discovered during their brief outbursts, which requires that the outburst occur at a time when it can be observed.  Since X-ray astronomy is only $\approx 50$ years old, objects that have not gone into outburst during that time cannot have been identified.  Many objects have been observed in outburst only once, which implies the existence of objects that haven't gone into outburst at all.  If such objects are common, the number of QBHBs in the galaxy will scale with the typical outburst recurrence time, a quantity for which there is currently no empirical constraint.

Theoretical arguments can be made regarding the outburst recurrence time based on assumptions associated with the disk instabilities responsible for the outbursts (e.g. \cite{WvP96}).  However the variable recurrence timescales displayed by objects that have multiple outbursts demonstrate that we still have much to learn about these instabilities.  But we now have decades of monitoring of QBHBs, and we can begin to see how they evolve as the disk builds up toward its next outburst.  Recently, a decade-long brightening trend has been found in Nova Muscae 1991 (\cite{W16}).  It is also notable that the passive quiescent state in A0620 (\cite{C10}) has disappeared over the past ten years in favor of the brighter active quiescent state.  We therefore examined all of the quiescent data from A0620 accumulated since 2003.  As can be seen in Figure 3, the initial finding was very encouraging --- however when data from 2015-16 were included, the inferred increase diminishes dramatically (although it is still significantly $>0$). As in many situations when studying quasi-periodic phenomena, it is important to beware mistaking short term trends for periodic cycles or long-term evolution.

\end{document}